# Magnetic Field Magnitude Modification for a Force-free Magnetic Cloud Model

R. P. Lepping[1,2] • C.-C. Wu[3] • D. B. Berdichevsky[4] • C. Kay[5]

[1] UMBC, Baltimore, MD 21250, USA
e-mail: Ronald.P.Lepping@gmail.com

[2] Heliophysics Science Division,
NASA/Goddard Space Flight Center,
Greenbelt, MD 20771, USA

[3] Naval Research Laboratory
Washington, DC 20735, USA

[4] IFIR/UNR-CONICET, 2000 Rosario, Sta Fé, Argentina, and
Code 672, GSFC/NASA, Greenbelt MD 20771, USA

[5] Solar Physics Division
NASA/Goddard Space Flight Center,
Greenbelt, MD 20771, USA

**Abstract**     A scheme was developed by Lepping, Berdichevsky, and Wu (*Solar Phys*., doi.10. 1007/ s11207-016-1040-9, 2017) [called the LBW article here] to approximate the average magnetic field magnitude ($B$-) profile of a "typical" magnetic cloud (MC) at/near 1AU.   It was based on actual *Wind* MC data, taken over 21 years, that was used to modify a time shifted Bessel function (force-free) magnetic field, where shifted refers to a field that was adjusted for typical MC self-similar expansion.   This was developed in the context of the Lepping, Jones, and Burlaga (*J. Geophys. Res,* vol. 95, p.11957, 1990) [called LJB here] MC parameter fitting model and should provide more realistic future representations of the MC's $B$-profile in most cases.   In the LBW article, through testing, it was shown that on average it can be expected that in about 80% of the MC cases (but it varies according to the spacecrafts' actual closest approach distances) the modified model's $B$-profile of the MC should be significantly improved by use of this scheme.   We describe how this scheme can be employed practically in modifying the LJB MC fitting model, and we test a new and slightly better (and less unwieldy) version of the scheme, the non-shifted (of Bessel functions) version, which is the one actually used in the LJB model modification.   The new scheme is based on modification formulae that are slightly more accurate than the old scheme, and it is expected to improve the $B$-profile in approximately 83% of the cases on average.   The schemes are applicable for use with data originating only at/near 1 AU, since the magnetic field and plasma data used in the development of the associated formulae were from only the *Wind* spacecraft, which was and is at 1 AU.

## 1. Introduction and Purpose

The Lepping, Jones, and Burlaga (1990) magnetic cloud (MC) parameter fitting model (henceforth called the LJB model), which assumes, over the path of the observing spacecraft through the structure, a force-free flux rope of cylindrical geometry and constant cross-section, has generally enjoyed success for the least squares fitting of hundreds of MCs. Success here means finding acceptable values for most of the free parameters, at least for the better quality cases within the limitations of the assumptions. However, the model has been clearly inadequate for accurately reproducing the intensity (magnitude) of the observed magnetic field along the path of the spacecraft through the MC. We address that weakness here. The central formulation of the LJB model, especially the force-free assumption, employs a Bessel function magnetic field (see Lundquist, 1950), which has a symmetrical magnetic field magnitude ($|\boldsymbol{B}| = B$) with respect to distance ($U$, in percentage) of the spacecraft's path through the MC. However, the actual observations very often show $B$ starting distinctly higher than the Bessel function field value, so that the overall linear tendency of $B$ is from high to low values over $U$, as discussed by Lepping, Berdichevsky, and Wu (2017) (henceforth called LBW), although in such cases the actual peak in $B$ is not necessarily exactly at the entrance point ($U = 0\%$) of the MC. There are also unusual cases where the $B$-profile is somewhat flat or it may even increase. A few examples are useful.

See Figure 1 which shows three examples of MC $B$-profiles: the first two (Figure 1A,B) of a decreasing $B$-profile and a third (Figure 1C) of an infrequently occurring ("contrary") case of an increasing $B$-profile, to first order (which occurred on November 11, 2009). Each example also shows the ideal Bessel function magnitude (given as a black solid curve) computed for the proper estimated closest approach value, for comparison. The example in Figure 1A is qualitatively a highly probable case and, in fact, is based on a smoothed average of many actual MCs. Specifically it is based on an average of $B/B_0$ ($B_0$ = estimated axial field magnitude) for 55 good quality cases having closest approach distances between 0% and 25% of the MCs' radii, as described in LBW (see set A of their Figure 4). LBW speculates as to why this profile appears this way (*i.e.* high to low values of $B$) which includes the MC's interaction with the upstream solar wind and its usual self-similar expansion, which occurs at least out to 1 AU and possibly much farther. Such effects are known to cause a field intensity enhancement in the early part of the MC. Figure 1B shows the MC (occurring on May 15, 2005) with the strongest estimated axial field within the *Wind* set (up to this point), and it was moving at about 843 km s$^{-1}$, on average. So when compared to a typical solar wind speed of 420 km s$^{-1}$, this MC was moving very fast causing considerable front-side compression and providing the severely enhanced $B$ at the front of the structure; its estimated axial field strength was $B_0 = 70.6$ nT. By contrast, Figure 1C's case was apparently caused by fast plasma impinging on the rear of the MC causing an enhancement in $B$ at the rear and, therefore, gives a $B$-profile that shows the uncommon low-to-high values. Figure 1A includes (in its average) a small number of such contrary profiles as that given in 1C, and that partly explains why its $B$-profile does not drop off as sharply as the specific profile in Figure 1B even when rendered in the normalized form $B/B_0$. Also, we point out that, because of the very large value of $B_0$ (= 70.6 nT) in panel B, if we had expressed $B$ without the $B_0$- normalization *vs.* %-duration, the red dotted curve would have been very steep, but would be difficult to compare to the curves in the other panels, A and C.   In order to get a broader understanding of how $B/B_0$ is ideally expected (here meaning according to the force-free model) to vary with respect to percent duration of travel through a MC parameterized by the spacecraft's closest approach value, CA, we show Figure 2; also see Figure 3 which defines CA in a cloud coordinate system, *i.e.* in terms of CA $\equiv Y_0/R_0$, where $Y_0$ is the spacecraft's estimated closest approach distance and $R_0$ is the MCs estimated radius.   The differences in the curves $B/B_0$ in Figure 2 are most dramatic at %-duration = 50%, of course. We should keep in mind, however, that the actual time spent within the MC for CA values near 0.95 are much shorter than those near CA = 0.1.   As Figure 1 shows, real observations of the $B$-profiles rarely show these ideal symmetric forms.

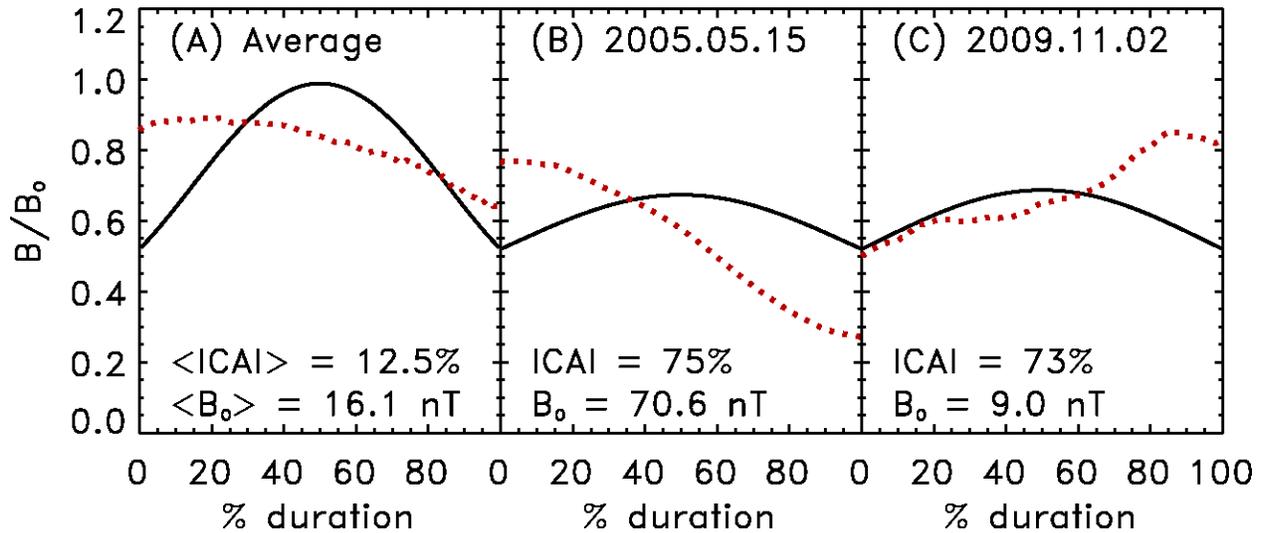

**Figure 1** Three examples of the differences between an actual magnetic field magnitude ($B/B_0$ red dotted curves) and a model field magnitude (solid curves) based on Bessel function fields, resulting from the LJB scheme: (A) Typical average profile of $B/B_0$ showing a generally decreasing *vs.* percent duration of path through the MC (from Figure 4 of LBW, curve A), (B) Extreme example (the May 15, 2005 case) of a steep decrease of $B/B_0$, and ( C) A contrary case (MC of Nov. 2, 2009) showing a $B/B_0$-profile rising in time across the MC, to first order (a low probability case).   All curves are normalized by the estimated magnetic field magnitude at the MC's axis, $B_0$, and in the case of panel A the $B_0$ is for each separate MC that went into the average (see curve A of Figure 4 in LBW), where $<B_0>$ = 16.1 nT, is only a representative value in that it is the average of the estimated $B_0$s for the full mission.   The applicable /CA/ and $B_0$ are presented in the lower portion of each panel.

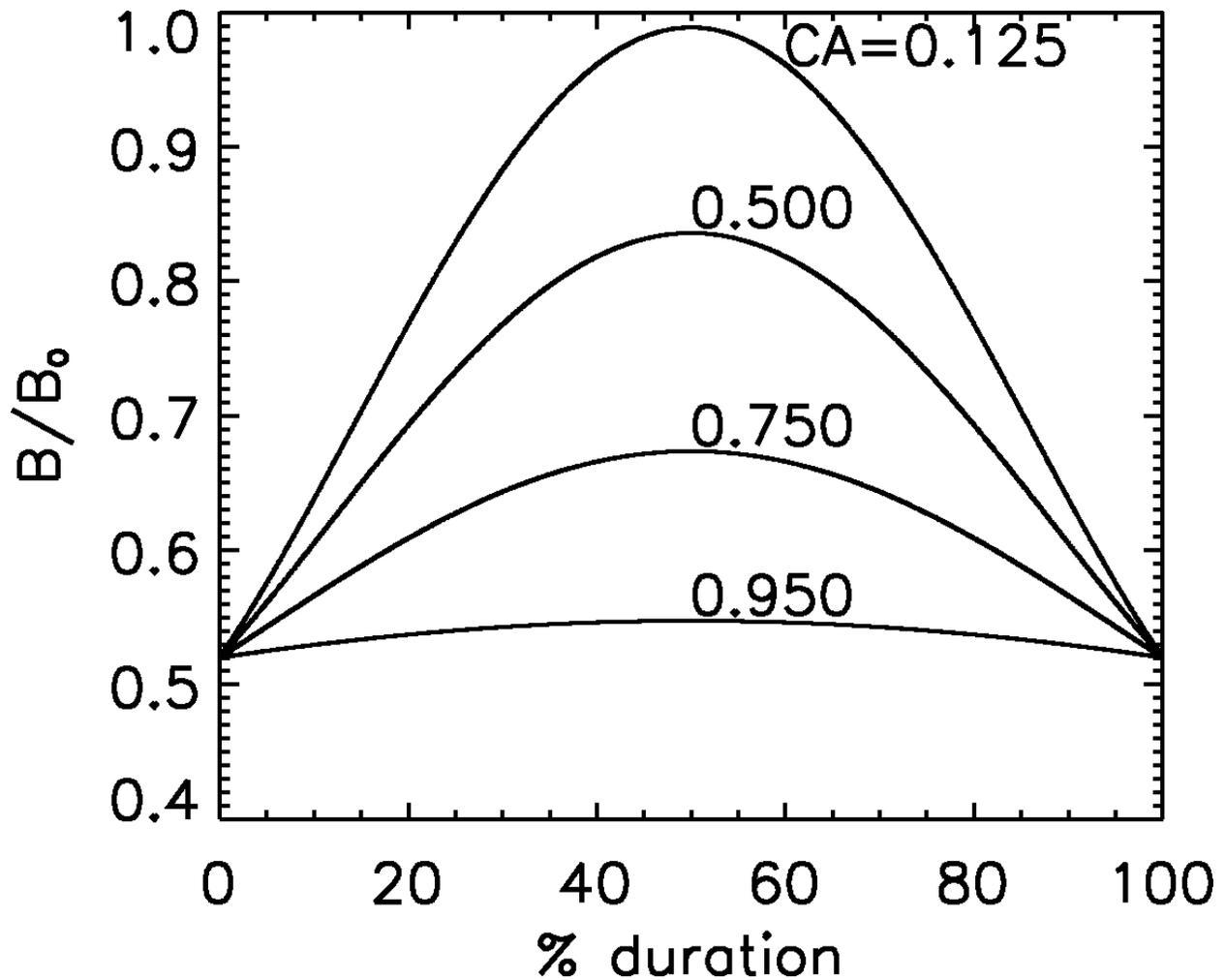

**Figure** 2 A family of plots of *B/B₀* *vs.* percent duration of the spacecraft through a MC based purely on Bessel function values, as expressed by Equation 2 for four values of CA as labeled on the curves from a small CA (= 0.125) to a very distant CA (= 0.95), and two intermediate values (CA = 0.50 and 0.75). *B/B₀* has the same value (0.520) for all endpoints which is reasonable, because at the endpoints the MC's field is ideally all tangential, where $\alpha R = \alpha R_0 = 2.40$ in Equation 2, a fixed value.

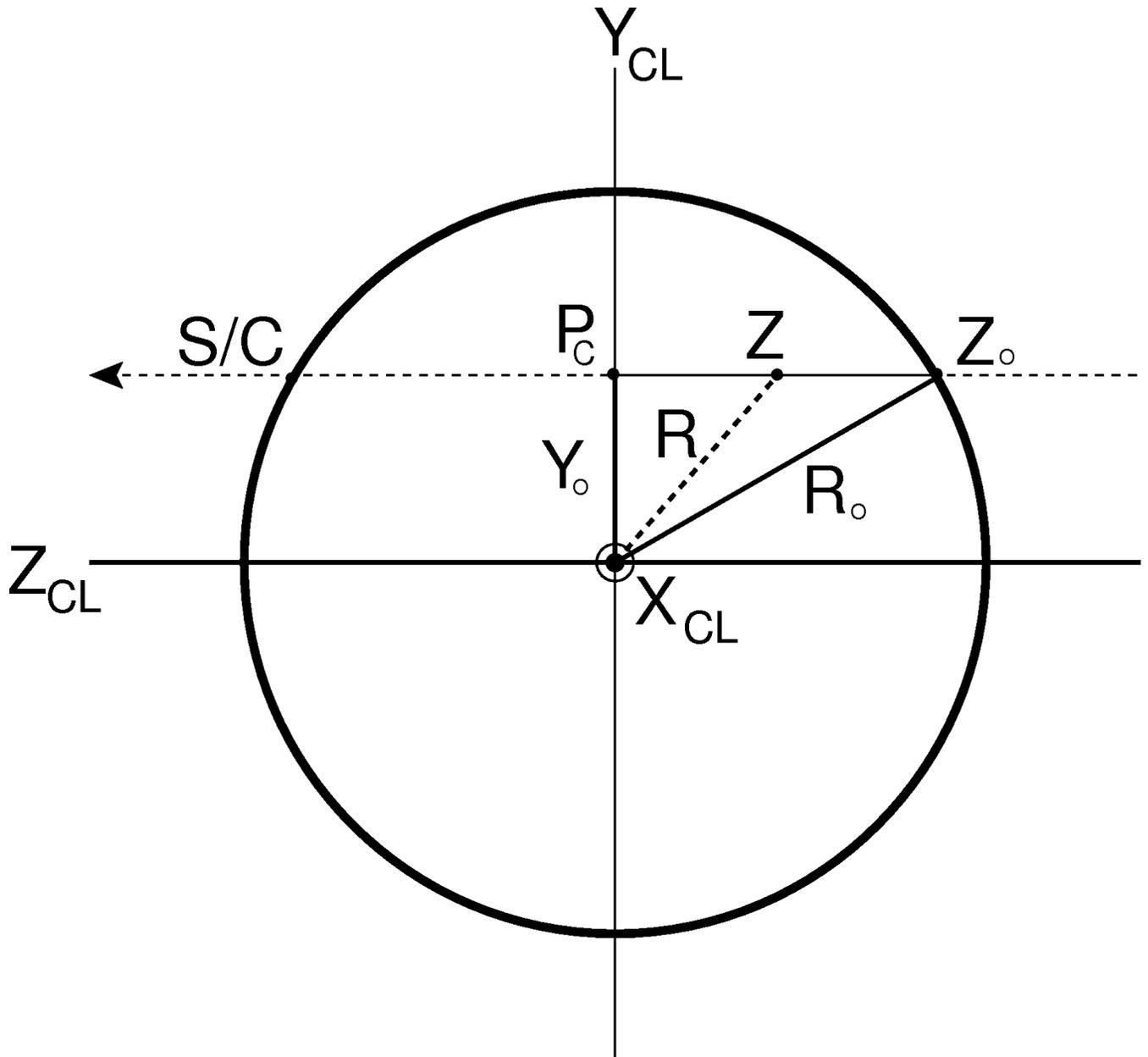

**Figure 3** Representation of the MC's cross-section in MC (Cl) coordinates. The projection of the spacecraft's path onto the cross-section defines the $\mathbf{Z_{Cl}}$-axis in the Cl system, and $Y_0$ is the closest approach distance. Note that $\mathbf{Y_{Cl}} = \mathbf{Z_{Cl}} \times \mathbf{X_{Cl}}$ in the Cl system, where $\mathbf{X_{Cl}}$ is aligned with the axis of the MC (and positive in the direction of the axial magnetic field), and $Y_0$ is along $\mathbf{Y_{Cl}}$.

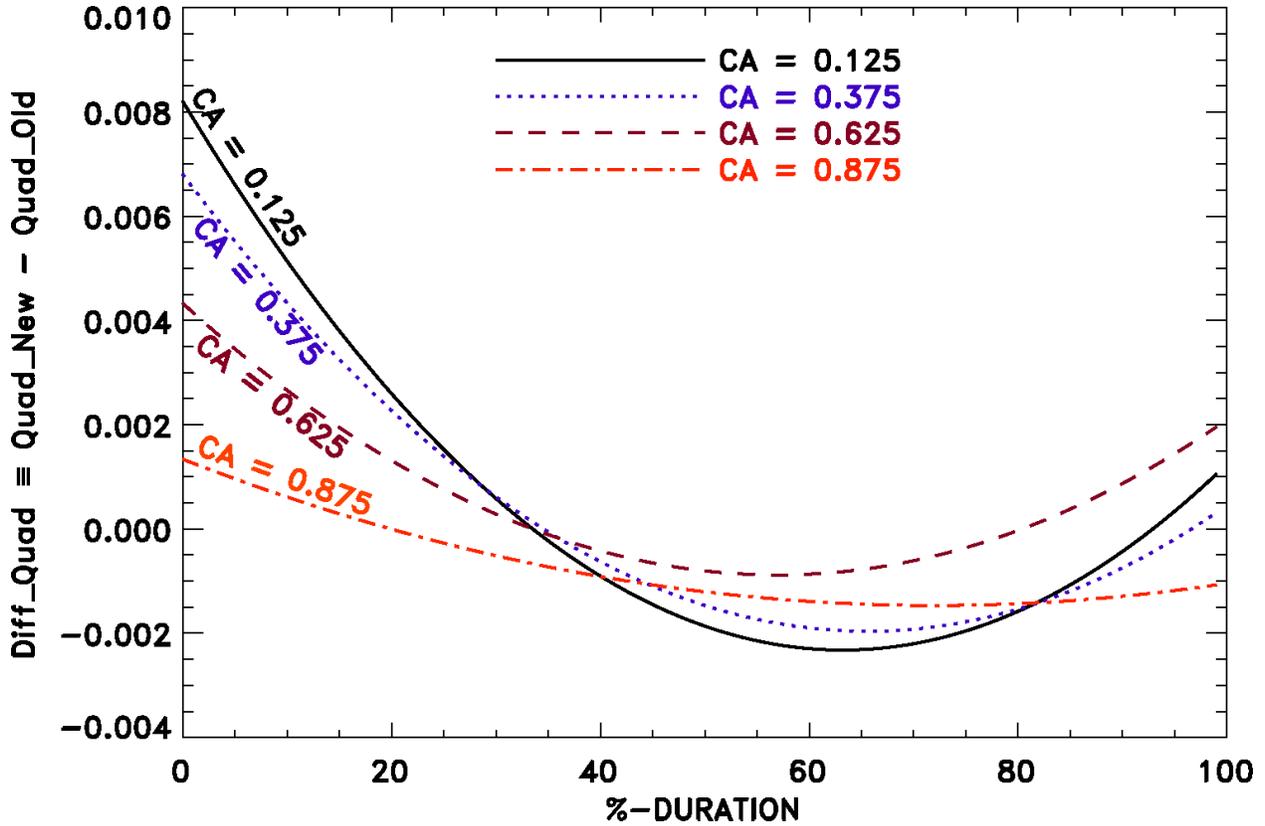

**Figure 4.** Plot of the Diff_Quad ≡ Quad_New - Quad_Old field derived from two different types of the Bessel functions, *i.e.* shifted (for the old scheme) and non-shifted (for the new scheme); see the text.

We point out that in the LJB model the observed magnetic field is at first unit-normalized at each measured point, then least-squares fitted to a unit-normalized Bessel function field, and at the end of the process made to satisfy the average field magnitude across the full MC where the estimated closest approach distance is taken into account. In this way, the magnetic field directional change of each measured point is given precedence over field magnitude change which is handled in a later step, in an average way, in the model. So consistent with the original LJB model, we take the same approach of treating the field magnitude separately in the magnitude-modification represented by this work.

Examples of other MC fitting models are: Vandas, Fisher, Geranios (1991), Marubashi (1997), Mulligan, Russell, and Luhmann (1998), Shimazu and Marubashi (2000), and Hidalgo, Nieves-Chinchilla, and Cid (2002). For other relevant references to the various studies of MCs in general over many years see the reference lists in LBW and in Lepping *et al.* (2006). The first to discover MCs in the solar wind was L. Burlaga in the late 1970s (see Burlaga *et al.*, 1980, 1981, and Burlaga, 1988, 1995); for an excellent tutorial on magnetic flux ropes/tubes see Priest (1990). MCs have been detected by many other spacecraft over the years (besides *Wind*), *e.g.* see Jian *et al.* (2018) for a recent study that examines *STEREO A/B* observations of ICMEs/MCs for the years 2007 - 2016, where ICME refers to interplanetary coronal mass ejection. Other such spacecraft are, for example, *IMP-8*, *Geotail*, *Ulysses*, *ACE*, *Voyager 1,2*, *Messenger*, *Helios A,B*, and others. For an example of case studies using multiple spacecraft data sets (*Voyager* and *Helios A,B*) see Berdichevsky *et al.* (2009), and for a *Ulysses* study, in which a multi-tube MC is observed at 4.69 AU, and where the authors argue that the force-free assumption is difficult to justify, see Osherovich *et al.* (1999); in the latter study implications suggest that complex MCs can exist in which multiple helical magnetic fields are embedded in cylindrically symmetric flux ropes. Global time-varying models of CMEs/ICMEs with embedded field structures that are usually MCs (or MC-like structures) are also developed, and in particular see Kay *et al.*(2017) for an example of such a model which addresses predicting the interaction of an ICME with the Earth's magnetosphere with specific examples.

In LBW the authors develop a *B*-modification equation, based on *Wind* MC data at 1 AU on average, that adjusts for a (proportionally) time-shifted Bessel function field, where shifted refers to using a Bessel function field that was adjusted for typical MC self-similar expansion, and where $\Delta V$ of 40 km/s was used for the average speed-difference across the MC (*i.e.* the speed from the time of entrance to that of exit of the MC) (*e.g*, see Lepping *et al.*, 2003). The resulting adjustment equation is a quadratic function of distance from the observing spacecraft's entry point of the MC. It is called the Quad(CA,*U*) formula, representing a correction to the LJB model's representation of <*B*/$B_0$> as a function of relative closest approach (CA) and *U*, where again CA $\equiv Y_0/R_0$, $Y_0$ is the spacecraft's estimated closest approach distance and $R_0$ is the MCs estimated radius (see Figure 3). This equation was obtained from a least-squares fit to a curve based on the difference (Diff) between the LJB model(shifted) and 101 averages of observed *B*/$B_0$ (across the average *Wind* MC profile of many events), *i.e.* averaged over many actual MCs (details below), and $B_0$ is the MC's estimated magnetic field magnitude on the MC's axis (from the LJB model) separately for each MC. Hence, we see that

$$\text{Quad(CA,}U) \approx <B/B_0> \text{ - Model(shifted)}, \tag{1}$$

where the approximation sign that replaces "=" is the result of replacing the Diff with the quadratic fit to obtain the Quad(CA,*U*), but as LBW argues, the approximation is a very good one, as Table 1 shows (*i.e.* small SigmaOs under the Old scheme). Equation 1 here is Equation 23 in LBW, and Model(shifted) [$\equiv B_M/B_0$(shifted)] is associated with Equation 3 from that same article, *i.e.*

$$B_M/B_0 = [(J_0(\alpha R))^2 + (J_1(\alpha R))^2]^{1/2}, \tag{2}$$

where $J_0$ and $J_1$ are the zeroth and first order Bessel functions, respectively, and $\alpha = 2.40/R_0$ [the so-called Lundquist (1950) solution], unless it is a core-annulus case, which is discussed by Lepping *et al.* (2006). [Equation 2 does not explicitly show the time-shifting, nor do the following three equations.] In our case $\alpha R$ is given by

$$\alpha R = \alpha R_0 \, [(CA)^2 + (Z/R_0)^2]^{1/2}, \tag{3}$$

**Table 1.** Coefficients for the Old and New Quad Formulae versus CA.

| Label | CA[a] | | Old Scheme[b] | | | | New Scheme[c] | | | |
|---|---|---|---|---|---|---|---|---|---|---|
| | $(N_{MC})$[d] | (% of $R_0$) | $co_1$ | $co_2$ | $co_3$ | $\sigma_0$ | $cn_1$ | $cn_2$ | $cn_3$ | $\sigma_0$ |
| A | (55) | 12.5 | 0.364 | -0.0174 | 0.000155 | 0.0173 | 0.372 | -0.0178 | 0.000158 | 0.0174 |
| B | (39) | 37.5 | 0.291 | -0.0131 | 0.000110 | 0.0129 | 0.298 | -0.0133 | 0.000112 | 0.0127 |
| C | (18) | 62.5 | 0.219 | -0.0096 | 0.000078 | 0.0234 | 0.224 | -0.0097 | 0.000080 | 0.0155 |
| D | (12) | 87.5 | 0.141 | -0.0050 | 0.000024 | 0.0110 | 0.142 | -0.0051 | 0.000025 | 0.0110 |

[a] The value of CA here is the center point (in terms of percentage of $R_0$) of sector: A, B, C, or D.
[b] The old set of coefficients was first given in Table 2 of LBW, where Quad_Old = $co_1 + co_2\,U + co_3\,U^2$, and $U$ is in percentage of path traveled through the MC.
[c] Quad_New = $cn_1 + cn_2\,U + cn_3\,U^2$.
[d] $N_{MC}$ is the number of MCs that was used in obtaining the average $<|\mathbf{B}|>$ used to develop the Quad formulae, one for a each sector, but restricted to MCs of $Q_0 = 1,2$ only.

where $Z$ is the distance along the $\mathbf{Z_{CI}}$-axis in magnetic cloud (Cl) coordinates (see LBW and Figure 3); see Appendix A, which defines magnetic cloud (CL) coordinates.   Equation 3 can be given as

$$\alpha R = 2.40[(CA)^2 + u^2(1 - (CA)^2)]^{1/2}, \qquad (4)$$

where $|u|\ (\equiv |Z/Z_O|$, and where $Z_O$ is on the MC's boundary) is a relative path length of the observing spacecraft through the MC, as shown by Equation 9 in LBW.   Note that the range of $u$ is 0.0 to +1.0, and when thought of as percent of duration it is given by

$$U(\% \ ) = (Z/Z_0 + 1.0) \times 50\%, \qquad (5)$$

where $Z/Z_0\ = -1.0$ to +1.0, and then $U(\%) = 0\%$ to 100%, respectively.   We remind the reader that $u$ (and therefore $U(\%)$) is measured along the $\mathbf{Z_{CI}}$-axis in Cl coordinates.   In Equation 4 when $u = 1.0$ (*i.e.* at the end point of the spacecraft's path within the MC) we see that  $\alpha R\ = 2.40$, which is independent of CA.   Hence, from Equation 2 for $u = 1.0$ (the end-point), $B_M/B_0$ will have a fixed value, *i.e.* it is independent of the value of CA, and similarly for the starting point.   This phenomenon is evident in Figure 2, and likewise for Figure 1, where for all three examples (*i.e.* for all three values of CA) the end point has the same value of $B_M/B_0$ (= 0.520). That the beginning and end points have the same value of $B_M/B_0$ is obviously consistent with the fact that the "Bessel function magnitude" (shown in Figure 1 without any time-shifting) is symmetric about $u = 0.5$.   However, for the actual observations $B/B_0$ is virtually never symmetric about $u = 0.5$.   Finally, we again stress that Equations 2 through 5 are shown before the proportional field time-shifting, which is done separately.

Going "backward" and applying Equation 1 to any single given MC for correction of the magnetic field magnitude, we can write it as

$$B(\text{est})/B_0 \ \approx [\text{Model(shifted)} + \text{Quad(CA},U)], \qquad (6)$$

**Table 2** Values of sigma quantities used in the two $B/B_0$- correction schemes[a].

| Code | $\sigma_1$ | $\sigma_2$ | $\sigma_{N1}$ | $\sigma_{N2}$ | $\Delta\sigma$ | $\Delta\sigma_N$ | $\Delta\sigma_1$ | $\Delta\sigma_2$ | $\Delta\sigma_N/\sigma_{N2}$ |
|------|-----------|-----------|--------------|--------------|---------------|-----------------|-----------------|-----------------|------------------------------|
| E01 | 0.154 | 0.116 | 0.161 | 0.119 | 0.038 | 0.042 | 0.007 | 0.003 | 0.35 |
| E02 | 0.192 | 0.184 | 0.186 | 0.163 | 0.008 | 0.023 | -0.006 | -0.021 | 0.14 |
| E03 | 0.177 | 0.106 | 0.192 | 0.114 | 0.071 | 0.078 | 0.015 | 0.008 | 0.68 |
| E04 | 0.211 | 0.206 | 0.199 | 0.182 | 0.005 | 0.017 | -0.012 | -0.024 | 0.09 |
| E05 | 0.087 | 0.066 | 0.090 | 0.057 | 0.021 | 0.033 | 0.003 | -0.009 | 0.58 |
| E06 | 0.134 | 0.100 | 0.156 | 0.119 | 0.034 | 0.037 | 0.022 | 0.019 | 0.31 |
| E07 | 0.158 | 0.094 | 0.162 | 0.092 | 0.064 | 0.070 | 0.004 | -0.002 | 0.76 |
| E08 | 0.168 | 0.094 | 0.186 | 0.108 | 0.074 | 0.078 | 0.018 | 0.014 | 0.72 |
| E09 | 0.072 | 0.075 | 0.084 | 0.072 | -0.003 | 0.012 | 0.012 | -0.003 | 0.17 |
| E10[b] | 0.124 | 0.201 | 0.123 | 0.187 | -0.077 | -0.064 | -0.001 | -0.014 | -0.34 |
| E11[c] | 0.147 | 0.193 | 0.141 | 0.176 | -0.046 | -0.035 | -0.006 | -0.017 | -0.20 |
| E12 | 0.156 | 0.109 | 0.161 | 0.109 | 0.047 | 0.052 | 0.005 | 0.000 | 0.48 |
| E13 | 0.160 | 0.109 | 0.163 | 0.128 | 0.051 | 0.035 | 0.003 | 0.019 | 0.27 |
| E14 | 0.191 | 0.128 | 0.195 | 0.141 | 0.063 | 0.054 | 0.004 | 0.013 | 0.38 |
| E15 | 0.155 | 0.085 | 0.156 | 0.098 | 0.070 | 0.058 | 0.001 | 0.013 | 0.59 |
| E16 | 0.187 | 0.108 | 0.188 | 0.122 | 0.079 | 0.066 | 0.001 | 0.014 | 0.54 |
| E17 | 0.123 | 0.058 | 0.137 | 0.070 | 0.065 | 0.067 | 0.014 | 0.012 | 0.96 |
| E18 | 0.158 | 0.110 | 0.166 | 0.113 | 0.048 | 0.053 | 0.008 | 0.003 | 0.47 |
| E19 | 0.077 | 0.082 | 0.079 | 0.066 | -0.005 | 0.013 | 0.002 | -0.016 | 0.20 |
| E20 | 0.106 | 0.069 | 0.123 | 0.074 | 0.037 | 0.049 | 0.017 | 0.005 | 0.66 |

[a] The Old Scheme uses the shifted Bessel function model and the Old Quad (see LBW), and the New Scheme uses the non-shifted Bessel function and the New Quad. See Table 3 and Figure 8 of LBW for the original results using the Old Scheme for the same 20 events with the same Code numbers. In that table the date and start/end times are given for each event. Note that $\Delta\sigma \equiv \sigma_1 - \sigma_2$ and $\Delta\sigma_N \equiv \sigma_{N1} - \sigma_{N2}$ (see Equation 10), and

$\Delta\sigma_1 = \sigma_{N1} - \sigma_1$ and $\Delta\sigma_2 = \sigma_{N2} - \sigma_2$ (see Equations 13 and 14), where:

$\sigma_1 = \sigma$(Obs - Model(shifted)); see Equation 25 of LBW,

$\sigma_2 = \sigma$(Diff_Old - Quad_Old); see Equation 26 of LBW,

$\sigma_{N1} = \sigma_N$(Obs - Model); see Equation 8, where it is understood that Model is not shifted, and

$\sigma_{N2} = \sigma_N$(Diff$_N$ - Quad_New); see Equation 9.

[b] This is an unusual case where CA = 99% and also because $B$ increases across the MC, compared to most other MCs where $B$ decreases.

[c] This case is unusual in that $B$ markedly increases across the MC.

where we are generalizing from the average "MC" to predicting any specific case. This scheme's success rate strictly depends on the justification of that generalization and must be tested on past MCs. As we see, Quad(CA, $U$) in Equation 6 is a magnetic field magnitude modification to the LJB model, after the Bessel functions are shifted because of MC expansion. Such Quad formulae were developed for four closest approach (CA) values in LBW. The four equations were developed to represent averages for the four sectors, CA: 0.0 to 0.25 (called sector A), > 0.25 to 0.50 (sector B), > 0.50 to 0.75 (sector C), and > 0.75 to 1.00 (sector D), which have central points (in terms of percent of $R_0$) at 12.5, 37.5, 62.5, and 87.5 %, respectively. [It should be stressed that Equation 6 is applicable only for use for MCs at 1 AU since the magnetic field and plasma data used in the development of the Quad formulae were from only the *Wind* spacecraft which was and is located at/near 1 AU. Also the Model in Equation 6 must be the LJB model, because that was the model used in the equation's development, specifically to obtain CA and $B_0$]. Table 1, in the portion (near the center) referred to as the Old Scheme, gives the three coefficients for each of the four Quad formulae for these four sectors. SigmaO values in Table 1 are the sigmas (where sigma = √variance) of the quadratic curve fits for this scheme. The coefficients and sigmas in Table 1were based on the statistics of the first 21 years of *Wind* MC data (where $N_{MC}$ = 209 MCs), but in the early part of the study was restricted to MCs of quality $Q_0$ = 1,2 only (see Appendix B for a definition of quality, $Q_0$, based on Appendix A of Lepping *et al.* (2006)), which provides $N_{1,2}$ = 124 cases. Cubic fits were also calculated and shown to have sigmas that were very close to those based on the quadratic fits, and therefore they are not a significant improvement over the quadratic fits. Hence, they will not be considered further in this work. The LBW article provides the sigmas associated with the cubic fits for comparison.

**Table 3**    Interpretations of the various $\sigma$s and $\Delta\sigma$s.

| The $\sigma$ quantity of interest (at left below) is a measure[a] of how well .... |
|---|
| $\sigma_1$ ....    the old model (*i.e.* the shifted[b] model) fits the observations of $B/B_0$. |
| $\sigma_2$ ....    the old Quad equations fit the difference-profile between the observations and the old model values, *i.e.* a measure of the "final state." |
| $\sigma_{N1}$ .... the new model (*i.e.* non shifted) fits the observations of $B/B_0$. |
| $\sigma_{N2}$ .... the new Quad equations fit the difference-profile between the observations and the new model values, *i.e.* a measure of the "final state." Also see Equation 17 and associated text. |

| The $\Delta\sigma$ quantity of interest (at left below) is a measure[a] of the improvement    .... |
|---|
| $\Delta\sigma$ ..... in the fit of $B/B_0$ by adding in the old Quad; $\Delta\sigma$ must be greater than or equal to 0.0 for a success. |
| $\Delta\sigma_N$ .....in the fit of $B/B_0$ by adding in the new Quad; $\Delta\sigma_N$ must be greater than or equal to 0.0 for a success. |
| $\Delta\sigma_1$ .... between the shifted and unshifted models before the Quad-correction; $\Delta\sigma_1$ is typically positive because shifting improves the fit on average before the Quad modification is made. |
| $\Delta\sigma_2$ .... in $\sigma_2$ (after the Quad is employed) in going from the old scheme to the new. (Notice, for example, that in Table 2 there are twelve positive and eight negative.) |
| |
| $\Delta\sigma_N/\sigma_{N2}$ is a relative measure of the usual improvement in the $B/B_0$ fit by using Quad weighted by the quality of the final fit, for the new model (*i.e.* non shifted). |

[a] For all quantities the term "measure" here means a quantitative (RMS) measure, as described in the footnotes to Table 2 and the related defining equations provided there.

[b] Shifted refers to using, in the model, a Bessel function magnetic field that was time- adjusted for typical MC self-similar expansion, and where $\Delta V$ of 40 km/s was used for the average speed-difference across the MC (*i.e.* the speed from the time of entrance to that of exit of the MC) (*e.g*, see Lepping *et al.*, 2003).

By applying the LJB MC fitting model to a given MC's field data, we will obtain $B_0$, CA, and the appropriate Model(shifted) for each case which gives the right side of Equation 6, after the appropriate Quad(CA,$U$) formula is used, and usually interpolation of two Quad formulae (developed for two different average CAs) is necessary to obtain a Quad that is appropriate for a specific MC.

The main purpose of this article is to describe the development of an improved program for providing a magnetic field magnitude modification for the LJB MC model via a modification of the Quad (correction) formulae (similar to Equation 6) first introduced in LBW.    We argue that a new set of correction formulae ("New Quad formula," Equation 7 in Section 2 below) provides a slightly superior scheme than the earlier ("Old Quad") scheme using Equation 6.    We discusses the principal differences in the two schemes and give a detailed argument as to why the new one is superior (Section 3).    Appropriate testing of the overall scheme (for the old and new forms) is done in Section 4.    We give some examples of the implementation of the new modification scheme (Section 5).    Also we expect that both sets of Quad formulae (old and new) will help to give insight into the actual variation of $B$ within a MC, on average, by their comparisons with $B$ observations for individual MCs, as well as to provide a practical tool for correcting $B$ in LJB MC model fitting (at least for cases at or near 1 AU), as stated above.    So the study has a dual purpose.    Section 6 gives a summary and general discussion.

## 2.    The New Quad Formulae

The work leading up to developing the Quad(CA,$U$) equations was carried out to gain greater understanding of how the actual average field magnitude (based on many actual good quality MCs, *i.e.* those of $Q_0 = 1,2$) varies along the spacecraft's path through a MC and how it differs from the shifted Bessel function field magnitude, both parameterized according to CA and examined within the framework of the LJB model.    However, as stated above, the Quad(CA,$U$) formulae have a practical application.    They are also useful for adjusting the estimated field magnitude profile that is obtained for any given MC when using the LJB model and the Quad(CA,$U$) adjustment, even though the adjustment is based on an average of many MCs, as discussed by LBW - it is a statistical correction.    The original derivation of the Quad formulae was based on using a difference between the actual averaged data and the shifted Bessel magnetic field as discussed in the Introduction.

We now develop another set of Quad formulae based on the difference between the actual averaged data and the Bessel magnetic field without any shifting of the Bessel function field; we refer to this set as the New Quad Set.    Then analogous to Equation 6 we have an estimation of $B$ (now denoted as $B_N$(est), for $B$(est)_New, to distinguish it from the estimate that was based on the Old Quad formulae from the recent LBW study discussed in the Introduction), *i.e.*

$$B_N(\text{est})/B_0 \approx [\text{Model} + \text{Quad(CA},U)\_\text{New}], \tag{7}$$

where "Model" here represents simply the Bessel magnetic field without any time shifting (see

Equation 2).    Table 1 gives the three coefficients for the four Quad formulae for the New Scheme for these four sectors (A, B, C, D) just as we did for the Old Quad Scheme; see the right portion in Table 1 referred to as the New Scheme.    Notice that the sigmas for the New set of formulae are very close to being the same as, or are better than, those for the Old set, and notice especially the sigmas for sector C, being 0.0234 for the Old set and 0.0155 for the New set, a distinct improvement.

## 3.    Principle differences between the Old and New Quad Formulae

Here we discuss some differences between the Old versus New Quads (and their developments) and why using the new scheme (*i.e.* with no time shift in the Bessel functions) is considered the better one.    (We refer to these as the old and new schemes.)    First, in obtaining the Old Quads we had to provide an average $\Delta V$ (defined as the difference between the plasma speeds at the entrance and exit points of the MC) of 40 km/s based on many *Wind* MCs, but it was after all only an average value, and the actual values are spread over a large spectrum, from 0.0 km/s to over a $\approx 100$ km/s.    Obviously the use of $\Delta V$ of 40 km/s (or of any fixed choice) introduces an error.    Although believed to be small, this error can be avoided by using the New Quads which do not depend on the parameter $\Delta V$ at all, but, to be clear, the New Quad formulae also depend on MC expansion indirectly and in an average way, and of course, there is no need to assume the *type* of expansion (self-similar, or not).    Second, implementing the new scheme by using the New Quads for the correction of $B$ is simply easier, because we avoid the extra step of having to do the time shifting of the Bessel functions.    Third, it is also important to realize that (as stated above) the sigmas for the quadratic fits for the New Scheme are very close to being the same as or are better than those for the Old Scheme (especially for sector C), as Table 1 shows. That alone argues strongly for use of the New Scheme.    In Figure 4 we show plots of the difference between the Old and New Quads (Diff_Quad ≡ Quad _New - Quad _ Old) as functions of $U$, *i.e.* as %-duration of the spacecraft's path through the MC, for the four sectors of CA: A, B, C, and D. The plot shows that, in any case, the difference between the old and new Quads is small for all four sectors, being almost negligible for the D sector.    Further, as will be seen in Section 4, in the examples and testing section, on average there is very little difference in the degree of correction of $B/B_0$ between the two schemes, with the new scheme being very slightly better.

## 4.    Examples and Testing of the Old and New Quads

LBW shows examples of the use of the Old Quad formulae for modification of $B$ for many examples and compare the results with what one obtains without the modification, in an attempt to test the old scheme.    They show that for a large percentage of cases (*i.e.* ~85 % from Table 5 of LBW) the modification is clearly worth doing for most cases. Now we test the new formulae, Equation 7, for specific MCs (or for large sets of specific MCs), and refer to the specific observed sets of $B/B_0$ as Obs and the LJB model field ratio $B/B_0$ as the Model (without a shift), and these are both functions of $U$ (%-duration).    Also, for these specific cases we refer to $\text{Diff}_N$ ≡ (Obs - Model), which is also a function of $U$; and recall that Model will depend on CA through Equation 4.    Quad_New is just the quadratic function of $U$ for the appropriate CA (obtained from the Model without a shift, but otherwise developed as described above (*i.e.* by Equations 1 through 6), and interpolated if necessary in practice.    We define values of $\sigma_N$, (again subscript N on $\sigma_N$ is for New), to be used in objective testing when comparing the profiles, as

$$\sigma_{N1} \equiv \sigma_N(\text{Obs - Model}) \quad = [(1/101)\ \Sigma(\text{Obs}(U) \quad - \quad \text{Model}(U))^2]^{1/2} \tag{8}$$

and

$$\sigma_{N2} \equiv \sigma_N(\text{Diff}_N - \text{Quad\_New}) \quad = [(1/101)\ \Sigma(\text{Diff}_N(U) - \text{Quad}(U)\_\text{New})^2]^{1/2} \tag{9}$$

where $\Sigma$ is a summation over 101 terms (*i.e.* including the two end points).    We define $\Delta\sigma_N$ as

$$\Delta\sigma_N = \sigma_{N1} - \sigma_{N2} \equiv \sigma_N(\text{Obs - Model}) - \sigma_N(\text{Diff}_N - \text{Quad\_New}). \tag{10}$$

So when $\sigma_N(\text{Diff}_N - \text{Quad\_New}) \leq \sigma_N(\text{Obs - Model})$ holds, *i.e.* when

$$\Delta\sigma_N \geq 0.0, \tag{11}$$

we say that we have a success, in the sense that the use of the Quad\_New formula improves the modeling of the *B*-profile (or there is effectively no change), and then use of Equation 7 for correcting $B/B_0$ is appropriate.    When we have

$$\Delta\sigma_N < 0.0, \tag{12}$$

we refer to the attempt at correction a failure.

Table 3 of LBW gives a list of 20 MCs that were examined using the shifted Bessel function (Old) model and the Old Quad formulae, and it lists whether the scheme was a success or not, provided by Code number for each event.    [Figure 8 of that paper shows the actual observed *B*-profiles (*i.e.* Obs), Diffs, Quads, etc. for these 20 MCs.]    The table also gives the date, start/end times, duration, Quality, CA, Sector letter (A,B,C,D), and any relevant notes for these events.    Table 2 here lists these same events by the same Code number (E1 through E20), and also it gives the results of the New Model and New Quad for a double comparison.    These results include the values of $\sigma_N$ and $\sigma$ (the latter representing the Old scheme), along with $\Delta\sigma_N$ and $\Delta\sigma$ for comparison, to check for success (positive or zero value) or failure (negative value). Notice that there are 16 successes out of 20 cases (80% success rate) for the old scheme, and 18 successes out of 20 (90% success rate) for the new scheme.    (Since 20 MCs represents a very small sample, we examine the full set of *Wind* MCs below.)    For the Old scheme the average of $\Delta\sigma$ for the 16 successful cases is 0.048, and for the New scheme the average of $\Delta\sigma_N$ for the 18 successful cases is 0.047, slightly better but almost the same.

Now we define two new quantities:

$$\Delta\sigma_1 = \sigma_{N1} - \sigma_1 \tag{13}$$

and

$$\Delta\sigma_2 = \sigma_{N2} - \sigma_2, \tag{14}$$

where $\sigma_1 = \sigma$(Obs - Model(shifted)) and $\sigma_2 = \sigma$(Diff_Old - Quad_Old); see Equations 25 and 26 of LBW, respectively.    We provide values for $\Delta\sigma_1$ and $\Delta\sigma_2$ for the same 20 cases in the last two columns of Table 2.    $\Delta\sigma_1$ compares $\sigma_1$ (before the Quad is employed) between the Old and New Schemes, and $\Delta\sigma_2$ compares $\sigma_2$ (after the Quad is employed) between the Old and New Schemes.

  Since the Quads are used to improve the estimates of $B/B_0$ for both schemes, we expect, on average, that $\Delta\sigma_2 < \Delta\sigma_1$, and hopefully $\Delta\sigma_2 << \Delta\sigma_1$; see Table 3.    First, we note that most values of $\Delta\sigma_1$ values in Table 2 are positive (*i.e.* 16 out of 20) meaning that typically $\sigma_{N1} > \sigma_1$ (giving positive $\Delta\sigma_1$s).      This is reasonable, because the New scheme does not correct for the Bessel function shift; recall that this shift was expected to give smaller $\sigma_1$s than a scheme without the shift.    The average of all 20 $\Delta\sigma_1$ (seventh column of Table 2, excluding the Code col.) is 0.0056, and the average of all 20 $\Delta\sigma_2$s is 0.00085, giving a ratio $<\Delta\sigma_1>/<\Delta\sigma_2> = 6.5$.    It is not surprising that $<\Delta\sigma_2>$ is so small, and that $<\Delta\sigma_1>$ is significantly larger than $<\Delta\sigma_2>$.    It simply means that the magnitude correction scheme is generally working.    See Table 3 for interpretations of the various $\sigma$s shown in Table 2.    Now we examine much larger sets of MCs, the *Wind* MCs from launch to the end of 2015, to significantly improve our statistics.

  Guided by LBW (where the restriction of $\Delta\sigma \geq 0.0$, representing a success, was arbitrarily relaxed slightly to $\Delta\sigma > $ -0.02), we here define a "relaxed" failure for the new scheme:

$$\Delta\sigma_N \ < -0.02, \tag{15}$$

instead of that of Equation 12, and provide the results of testing in the last column of Tables 4 and 5, along with the results for Equation 12 strictly, in the next to last column. (Notice that the last two columns of Tables 4 and 5 have "$N$"-subscripts in the column-heading indicating results of the new scheme.)    In Tables 4 and 5 the percentage of failures are provided as a function of increased CA as we go up the rows of the tables, where the difference in the tables being that Table 4 examines all MCs (with total $N_{MC} = 209$, for $Q_0 = 1,2,3$) from the beginning of the *Wind* mission to the end of 2015 and Table 5 (with total $N_{MC} = 124$) examines only the MCs with $Q_0 = 1,2$ from the same time period.    In columns 3 and 4 are the same quantities except they are for the Old Quad cases, which were shown in Tables 4 and 5 of LBW, and are given here for comparison (with the slight change from LBW in that we now consider "less than and equal to" ($\leq$) various CAs, instead of only "less than" (<) in the first column of the tables, creating only slight changes in the outcome).    Also given in Tables 4 and 5 are the number of MCs (in parentheses) that the percentage represents for a given line. We now point out some relevant features of these tables.

  First, since we increase the number of MCs as we go up the rows of Tables 4 and 5, we then improve the statistics, *i.e.* we decrease the uncertainty in the estimate of the number of failed cases, as described in the footnotes of the tables. For both Tables 4 and 5 we see that, in all cases (*i.e.* for all CA-ranges and for both failure criteria, $\Delta\sigma_N < 0.0$ and $\Delta\sigma_N < $ -0.02 here), the new scheme gives a smaller percentage of failed cases. The averages below each column indicate that, for appropriate comparisons of Tables 4 and 5 (*i.e.* the last column of Table 4 against last column of Table 5, etc.) there are little differences, but the last columns, with 12.2% versus 11.3%, show that the better quality MCs with the looser failure criteria give the better results on average, as expected. The "spreads" (at the bottom of the tables) are in a sense a measure of stability in our statistics, with respect to changes in CA; they are derived from examining the max and min values against the average for each column. Notice that the spreads are generally smaller for Table 5 (where $Q_0 = 1,2$) than for Table 4 ($Q_0 = 1,2,3$) or about the same, as might be expected.

**Table 4**   Examining Percentages of Failed Attempts[a] for Quality $Q_0 = 1,2,3$ for Quad_New versus Quad_Old.

| CA Range | $N_{MC}$ | % having $\Delta\sigma < 0.00$ (No.) | % having $\Delta\sigma < $ -0.02 (No.) | % having $\Delta\sigma_N < 0.00$ (No.) | % having $\Delta\sigma_N < $ -0.02 (No.) |
|---|---|---|---|---|---|
| All CA | 209 | 26.3 % (55) | 18.2 % (38) | 22.0 % (46) | 15.8 % (33) |
| $\leq 0.75$ | 184 | 23.4 % (43) | 15.2 % (28) | 19.0 % (35) | 12.5 % (23) |
| $\leq 0.50$ | 155 | 19.4 % (30) | 16.1 % (25) | 16.8 % (26) | 12.8 % (20) |
| $\leq 0.25$ | 90 | 15.5% (14) | 11.1 % (10) | 12.2 % (11) | 7.8 %    (7) |
| Average | | 21.2 % | 15.2 % | 17.5% | 12.2% |
| Spread | | ($\pm \approx 6\%$) | ($\pm \approx 4\%$) | ($\pm \approx 5\%$) | ($\pm \approx 4\%$) |

[a] For a success in the strict limit the requirement is $\Delta\sigma_N \geq 0.0$ (see Equations 8 through 11), but for the relaxed limit success is $\Delta\sigma_N \geq $ -0.02 and failure $\Delta\sigma_N < $ -0.02 (Equation 15). Hence, we are listing here the percentage of failures for both the restricted and relaxed cases.
The same type of inequalities hold for $\Delta\sigma$ (see LBW).

**Table 5**   Examining Percentages of Failed Attempts[a] for only Quality $Q_0 = 1,2$ for Quad_New versus Quad_Old.

| CA Range | $N_{MC}$ | % having $\Delta\sigma < 0.00$ (No.) | % having $\Delta\sigma < $ -0.02 (No.) | % having $\Delta\sigma_N < 0.00$ (No.) | % having $\Delta\sigma_N < $ -0.02 (No.) |
|---|---|---|---|---|---|
| All CA | 124 | 24.2 % (30) | 16.1 % (20) | 19.4 % (24) | 12.1 % (15) |
| $\leq 0.75$ | 114 | 23.7 % (27) | 15.8 % (18) | 19.3 % (22) | 11.4 % (13) |
| $\leq 0.50$ | 94 | 19.1 % (18) | 18.1 % (17) | 18.1 % (17) | 12.8 % (12) |
| $\leq 0.25$ | 56 | 16.1%    (9) | 14.3 % (8) | 14.3 %    (8) | 8.9 %    (5) |
| Average | | 20.8 % | 16.1 % | 17.8% | 11.3% |
| Spread | | ($\pm \approx 4\%$) | ($\pm \approx 2\%$) | ($\pm \approx 7\%$) | ($\pm \approx 2\%$) |

[a] For a success in the strict limit the requirement is $\Delta\sigma_N \geq 0.0$ (see Equations 8 through 11), but for the relaxed limit success is $\Delta\sigma_N \geq $ -0.02 and failure $\Delta\sigma_N < $ -0.02 (Equation 15). Hence, we are listing here the percentage of failures for both the restricted and relaxed cases. The same type of inequalities hold for $\Delta\sigma$ (see LBW).

However, when one examines the difference in the spreads on the basis of the old *vs.* new schemes, we see mixed results which is difficult to interpret.    For example, the loose criteria ($\Delta\sigma < $ -0.02 and $\Delta\sigma_N < $ -0.02 for failure) give the same spread (within a given table), for both tables.    However, for the standard criteria ($\Delta\sigma < 0.0$ and $\Delta\sigma_N < 0.0$ for failure) the better quality MCs do not always give the better results.    But the spread is probably the least important aspect of these tables.    Finally, we point out the remarkably low percentages of failures estimated for the MCs with smallest CAs (*i.e.* $\leq 0.25$): 7.8% for those with $Q_0 = 1,2,3$ MCs and 8.9% for those with for $Q_0 = 1,2$ - perhaps backwards according to common intuition, but, as pointed out above, the statistics are somewhat poor in both of these cases.    (Also we remind the reader that the assignment of a $Q_0$ value does not depend on $B/B_0$, as pointed out in the Introduction, so we probably should not expect any significant dependence of "success" here on $Q_0$.)    In the final analysis, we conclude that either scheme for correcting for a typical $B$-profile of a 1 AU MC that usually does not follow the simple Bessel function field of Equation 2 (*i.e.* Old Quad or New Quad scheme) is on average useful, and the new scheme is the better one for the reasons given Section 3 and further because of the results shown in Tables 4 and 5.

Now we show why $\sigma_{N2}$ is an absolute error of $B(\text{est})/B_0$.    Consider the definition of $\sigma_{N2}$, given by Equation 9, which is (where $\sigma_N$ on the right side is an operator):

$$\sigma_{N2} \equiv \sigma_N(\text{Diff}_N - \text{Quad\_New})$$

and $\text{Diff}_N$ is $\text{Diff}_N \equiv (\text{Obs} - \text{Model})$.    Hence, we have

$$\sigma_{N2} \equiv \sigma_N(\text{Obs} - \text{Model} - \text{Quad\_New}). \qquad (16)$$

Or this can be expressed (using the definition of $B_N(\text{est})/B_0$ from Equation 7) as

$$\sigma_{N2} \approx \sigma_N(\text{Obs} - B_N(\text{est})/B_0). \qquad (17)$$

Since $B_N(\text{est})/B_0$ is an estimate of what we are trying to estimate, Obs, then $\sigma_{N2}$ is an estimate of the error in that operation using this Quad\_New scheme.    The value of $\sigma_{N2}$ can be determined for each MC, as we do below for some examples.

## 5.    Examples of Implementation

In Figure 5 we show eight examples of the use of the Quad(new) formulae (see Equation 7) where the ordinary Bessel functions (Equation 2) are used for the Model (given by the black solid curves in the figure), and where the sum of the Model and Quad(new) provides $B_N(\text{est})/B_0$ (red dashed curves).    The latter are representations of the observed $B_N/B_0$ for these events which are shown as the dash-dot-dot-dot curves.    The eight panels give examples of various levels of success of the scheme: Figure 5A) gives four examples of cases of excellent/good success, and 5B) gives three cases of intermediate but acceptable success, and one case (bottom right, example E11) shows a failure.    All code numbers are related to Table 2. (Table 2 shows an even worse

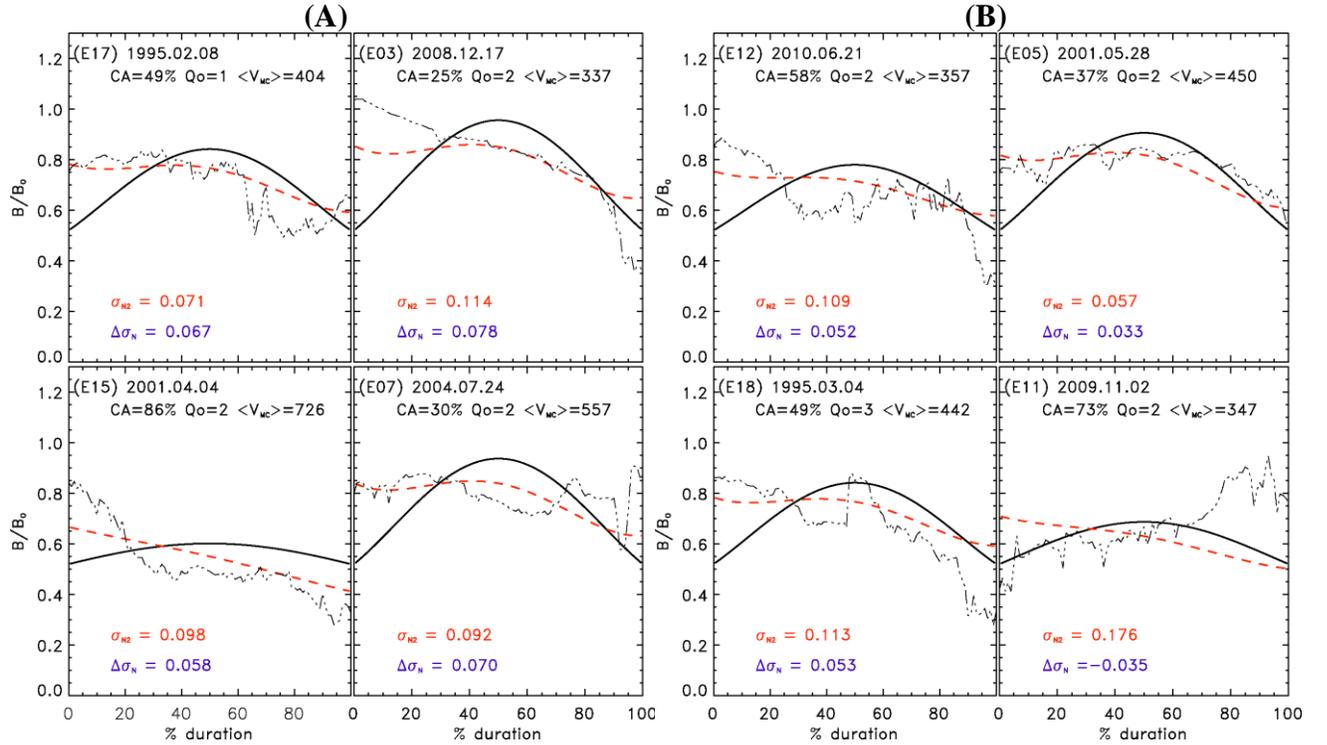

**(A)**                                           **(B)**

**Figure 5.** Examples of the use of the Quad_new formulae in Equation 7 where the ordinary Bessel functions (Equation 2) are used for the Model (given by the black solid curves), and where $B_N(\text{est})/B_0$ is shown as red dashed curves. The latter are representations of the observed $B_N/B_0$ (the dash-dot-dot-dot curves) for these events. The date of occurrence of each MC is given at the top of its panel. A total of eight panels are shown giving examples of various levels of success: (5A) Four examples of cases of excellent/good success, and (5B) Three cases of intermediate success, and one (bottom right, case E11) of a failure. All code numbers are from Table 2, from where the values for the $\sigma_{N2}$s and the $\Delta\sigma_N$s shown in the figure panels are also obtained. Recall that $\sigma_{N2}$ is an estimate of the absolute error in $B_N(\text{est})/B_0$ as an estimate of Obs (actual data) due to obtaining it from the New Quad scheme (see Equation 17).

failure than E11, *i.e.* case E10, but we do not show it as an example in Figure 5B, because it is an unusually poor case with an estimated CA of 99%; in other words, the actual CA value is very poorly know, but it is highly likely to be very large. Also, in this set of 20 MCs case E10 is the only case worse than E11, with respect to our $B_N/B_0$- modification scheme - see col. $\Delta\sigma_N$ of Table 2.) The code numbers relate originally to the earlier LBW article (see its Figure 8). Notice that almost all panels of Figure 5 are of quality $Q_0 = 2$, *i.e.* all except two cases: the top-left case (E17) of Figure 5A which has a quality of $Q_0 = 1$, and the bottom left case of Figure 5B (E18) which has a quality $Q_0 = 3$. The values of the related $\sigma_{N2}$s and the $\Delta\sigma_N$s for these examples (also shown in their panels) more-or-less tell the story of how the order of the success levels are chosen for these cases; they depend only on the magnitude profiles - not on quality, $Q_0$, which is based on various aspects of the LJB model fitting, *i.e.* mainly on field directional changes as the spacecraft travels through the MC. (However, it is no surprise that case E17, listed as first panel in Figure 5A, has a $Q_0 = 1$ and is an excellent $B_N/B_0$ correction.)    Along these same lines, notice in the bottom two panels of Figure 5B that case E18 with $Q_0 = 3$ is distinctly better than its neighbor E11 (last panel) which has $Q_0 = 2$.    Again, this can happen because estimation of quality is not dependent on field magnitude.

Typically the smaller $\sigma_{N2}$ is the better, and the larger $\Delta\sigma_N$ is the better, for field magnitude correction.    Hence, the ratio $\Delta\sigma_N/\sigma_{N2}$ ($\sigma_{N2} \neq 0.0$) is a very good indicator of success and, better still, it gives a quantitative level of success where high values (*i.e.* $\approx 1.0$) for this ratio indicate high levels of success.    (This is the reason example E17 was chosen as the first panel of Figure 5A, as we will see below.)    See the last column of Table 2 for evaluations of this ratio for the 20 events of LBW.    A smaller $\sigma_{N2}$ (which is an absolute error) represents good final "states" in an absolute sense (see Equations 9 and 17), and a larger $\Delta\sigma_N$ indicates that there was a large improvement (provided the value of $\Delta\sigma_N$ is positive), or at least a significant improvement compared to the initial state, where initial and final "state" here mean the $B_N/B_0$ profiles before and after the Quad-modification is used, respectively.    Notice that $\Delta\sigma_N$ was negative in example E11 of Figure 5B, showing that we made the situation worse by using the Quad-modification in this case, but all of the other seven examples shown in Figure 5 are improvements at various levels.    Notice in Table 2 that only two $\Delta\sigma_N$ values, out of 20, are negative; this is reasonable according to our more general statistical study whose results are given in Tables 4 and 5.    It is interesting that when examining the ratio $\Delta\sigma_N/\sigma_{N2}$ for the four cases of Figure 5A, we find that E17's ratio is approximately 1.0 (excellent) and its values for the other three cases are between $\approx$ 0.6 and $\approx$0.8 (both good), where obviously the higher the better.    Examining this same ratio for the examples of Figure 5B gives a very poor value of $\approx$ -0.2 for case E11, but the other three are between $\approx$0.5 and $\approx$0.6 (intermediate values but clear successes).

## 6.    Summary and Discussion

In LBW we calculated average $B/B_0$-profiles within MCs (at first using only cases of quality $Q_0 = 1,2$) at/near 1 AU based on 21 years of *Wind* magnetic field and plasma data; with this quality restriction this interval contained $N_{1,2} = 124$ MCs.    These average profiles were categorized into four evenly divided radial sectors depending on the values of estimated CA, which were from the LJB MC (force-free) fitting model.    This model also estimated the $B_0$ values for each MC, so that the values of $B/B_0$ were not fully model-free.    The four average $B/B_0$ -profiles from the $N_{1,2}$ = 124 MCs were compared to the proper $B/B_0$-solutions of the LJB model (*i.e.* time-shifted Bessel functions to account for average self-similar expansion and the proper CA), creating difference-fields which were shown to well approximate quadratic functions of distance; these four resulting quadratic equations were called the Quads    One of the main products of this process was Equation 1.    Through a generalization of Equation 1 we arrive at Equation 6, which is expected to well approximate $B/B_0$ for a typical MC at 1 AU most of the time, and tests were done to see specifically what that meant quantitatively.    In LBW testing of this scheme was carried out for sets of both $Q_0 =1,2$ cases only and the full $Q_0 =1,2,3$ cases (where $N_{MC} = 209$ MCs), and it was shown that on average we can expect that for between 82% and 89% of the cases (depending on the value of CA) the $B$-profile of the MC should be improved by use of this scheme when the full set $Q_0 = 1,2,3$ cases are considered; slight improvement to between 84% and 86% of the cases become successes when only $Q_0 = 1,2$ cases are used. So it appears that the Quad formulae could have been developed with the full set of $Q_0 = 1,2,3$ cases with little diminution of usefulness. When the scheme is employed in a non-real time mode the user has the option of using it when it actually improves the MC's *B*-profile (and determination of this will usually be obvious), or not using it, so that the *B*-profile is not made worse, which is expected in ~ 20% of the cases depending on the value of CA on average. All of these results were based on the shifted Bessel function condition, the old scheme.

We discuss how the Quad-modification scheme can actually be employed in general to improve the LJB MC fitting results, and the study describes and tests a slightly better (and less unwieldy) version of the scheme, the non-time-shifted (of Bessel functions) version; see Equation 7. This new scheme is the one we actually use in the modification of the model. See Section 3 which gives three arguments for the greater desirability of using this new scheme. Briefly, these arguments are: (1) In obtaining the Old Quads we had to provide an average $\Delta V$ (<$\Delta V$>, *i.e.* the average change in speed across a large sample of *Wind* MCs) of 40 km/s, but the actual values of $\Delta V$ are spread over a large spectrum, from 0.0 km/s to over a $\approx$100 km/s. So the use of any fixed choice for $\Delta V$ introduces an error, even if small. (2) Implementing the new scheme by using the New Quads for the correction of *B* is simply easier and less unwieldy, because we avoid the extra step of having to do the time shifting of the Bessel functions. (3) The sigmas for the quadratic fits for the New Scheme are very close to being the same as or are better than those for the old scheme (especially better for the sector centered on a CA of 62.5%).

It may, at first, appear counterintuitive to arrive at the conclusion that the New Quad is usually better than the Old Quad in-as-much-as the Old Quad accounts for a phenomenon that the New Quad does not (*i.e.* MC expansion), so we attempt here to argue why such intuition can be (and usually is) wrong in this kind of scheme. Attempting to adjust for MC expansion by using a typical but fixed $\Delta V$ (= 40 km s$^{-1}$) for all MCs in determining the associated Quad, at first, seems like the proper approach. But apparently by not adjusting for each MC's own $\Delta V$ (in both deriving the associated Quad and in the implementation of the scheme) we introduce an extra error for each MC, and a different error for each which apparently out-weights any benefit that the expansion-adjustment may provide, as our tests show. That is very likely why using the New Quad is statistically more successful than using the Old Quad, and why the belief that using the New Quad is, at first, counterintuitive. That is why concentrating on only the benefit and not the shortcomings of using an average $\Delta V$ for adjustment for expansion is a mistake (but we suggest it is not necessarily a large one). Each MC demands its own tailored $\Delta V$-adjustment which is obviously also what should be done for case studies. Our magnetic field magnitude correction aims for a simple modification, based on statistics.

Equation 7 provides the specific formula $\{B_N(\text{est})/B_0 \approx [\text{Model} + \text{Quad}(\text{CA}, U)\_\text{New}]\}$ for this new scheme. The new scheme was developed in almost the same way as the old scheme, except in the new scheme the Model's "Bessel function magnitude" is not time-shifted, and therefore its resulting Quad-formulae will be slightly different. Notice that, in both schemes, $B_N$(est)/$B_0$ consists of two terms, the first of which is purely theoretical (one of the versions of the Bessel functions) and the second is mainly empirical (described in LBW for the old scheme) and based on *Wind* magnetic field and plasma data, but obtained within the framework of the MC Model, to provide the estimated CA and $B_0$ needed to obtain the correct Quad formula.    In the actual fitting procedure this Quad-term addition is a module whose use is optional with an indicator telling whether the option has been used or not - and it will be the one associated with the new scheme, of course.    (In practice the appropriate Quad(CA,$U$) formula to use, when improvement is possible, is usually an interpolation between two Quad formulae, for two different CA-sectors, one above and one below the exact CA of interest.)    Obviously, from our various tests (*i.e.* from Tables 2, 4 and 5, and Figure 5) we know that whenever the observed field magnitude profile (after some reasonable smoothing) is flat, or even more obviously, when it shows a rising *B*-profile (see Figures 1C and 5B, panel E11), it is almost always better not to use this option.    The new scheme is expected to improve the *B*-profile in ~ 83% of the cases on average as examination of Tables 4 and 5 indicate, but the uncertainty on this for any given MC is rather large, especially depending on the value of CA.

It appears that the ratio $\Delta\sigma_N/\sigma_{N2}$ is an excellent quantitative measure of how successful the Quad_New scheme is in correcting $B/B_0$ for any given MC, as Figure 5 and its related text show, and the expected success is only weakly dependent on the quality ($Q_0$) of the event; see Equations 8, 9, and 10 for definitions of $\sigma_{N1}$, $\sigma_{N2}$, and $\Delta\sigma_N$, respectively, and the last column of Table 2.    Smaller $\sigma_{N2}$s represent good final states in an absolute sense, and larger $\Delta\sigma_N$s indicate that there was a significant improvement, provided the value of $\Delta\sigma_N$ is positive; a negative $\Delta\sigma_N$ is an outright failure of the scheme.    In fact, the quantity $\sigma_{N2}$ was shown (see Equation 17) to be an absolute error in calculating $B$(est)/$B_0$ as an estimate of Obs (actual data) using the New Quad scheme.

It should be stressed that the principal Equations 6 and 7 are applicable for use with data originating only at/near 1 AU, since the magnetic field and plasma data used in the development of the Quad formulae were from only the *Wind* spacecraft, which was and is at 1 AU, and the Model in those equations must be the LJB force-free model, because that was the model used in the development of those equations.    The dependence on the model is explicit through the use of the model-estimated $B_0$s in normalizing the field of each MC before the $N = 124$ cases (*i.e.* with $Q_0 = 1,2$ only) were averaged to generate the associated Quad in the first place, done for the separated CA-sectors A, B, C, and D, which were in turn chosen by the model-estimated CAs. It is important to realize that once the Quad correction term is added to either the old (shifted) or new LJB (non-shifted) model, it can no longer be considered exactly force-free.

We have indicated that MC expansion often leads to non-symmetric MC magnetic field magnitude profiles and account for it in the Old Quad scheme.    However, it should be pointed out that magnetic erosion due to the interaction of the MC with the solar wind via magnetic reconnection may be another cause for such asymmetry (*e.g.*, see Dasso *et al.* 2006; Ruffenach *et al.*, 2015).    Since we prefer using the New Quad scheme where this asymmetry, regardless of its cause, is essentially accounted for by the associated New Quad formulae, the reason for the asymmetry is not important to the $B/B_0$-modification.

A recent comprehensive study of the internal magnetic field configurations of ICMEs also using *Wind* data was carried out by Nieves-Chinchilla *et al.* (2018) over essentially the same period as our study. [Their study also considers a more complex set of magnetic configurations called "magnetic obstacles" (MOs).] They find 337 cases of ICMEs from which 298 cases had well organized magnetic field topologies. For comparison, we identified $N_{MC} = 209$ MCs over essentially the same period also using *Wind* data. Since MCs are relatively well organized magnetic field configurations (and usually much more restrictive than ICMEs by definition), and therefore more-or-less represent a subset of the ICMEs, it is reasonable that our set should be smaller in number. Our number of "events" greatly increases when we consider the sum of MCs and MC-like objects (combined set), which are identified in a similar manner to MCs but are shown not to have quite the same organized helical field structure and therefore not able to be fitted by the LJB model. The class of events that Nieves-Chinchilla *et al.* (2018) call MOs is related, even to some extent by definition but in a somewhat complicated way, to our combined set. They correctly state that "The configuration of the magnetic field embedded in ICMEs is subject to considerable uncertainties." They make significant progress in exposing the broad range of characteristics of ICMEs with respect to their embedded magnetic fields, especially in the area of asymmetries, and in clarifying the deep complexity of the subject. In particular, the Nieves-Chinchilla *et al.* (2018) findings with regard to asymmetries in the magnetic field magnitude have similarities to what our Quad formulae are attempting to adjust.

The Nieves-Chinchilla *et al.* (2018) study is beyond the scope of our study here, but we support their point of view of how complex ICME structures are by at least pointing out the broad range of relative MCs/ICMEs occurrence rates found by various research teams. For example, the Nieves-Chinchilla *et al.* (2018) relative occurrence rate MCs/ICMEs is (209/337 =) 62%, using our value of $N_{MC} = 209$ MCs. By comparison, the occurrence of MCs/ICMEs was 25% from Cane and Richardson (2003), 28% from Wu and Lepping (2011), 41% from Bothmer and Schwenn (1996), and 30% from Gosling (1990). And we show that even in the broad range of possible $B/B_0$-profiles (recall it is purely a scalar quantity) there is obvious complexity in that the Quad-technique works for a large percentage of cases, but can fail quite badly for some events, perhaps for 15 - 20% of the cases, at 1 AU, as Tables 2, 4, and 5 show.

**Acknowledgments**    We thank the *Wind* MFI and SWE teams for the care they employ in producing the magnetic field and plasma data used in this study. C.K.'s research was supported by an appointment to the NASA Postdoctoral Program at NASA GSFC, administered by the Universities Space Research Association under contract with NASA, and by the Dept. of Physics, The Catholic University of America, Washington DC 20064, USA. This study was partially supported by the Chief of Naval Research, and NASA LWS program, Grant No. 80HQTR18T0023 (CCW).

**Conflict of interest**    The authors indicate that they have no conflicts of interest.

**Appendix A      Magnetic Cloud (Cl) Coordinate System**

In the Cl system the $\mathbf{X_{Cl}}$-axis is along the MC's axis, positive in the direction of the positive polarity of the axial magnetic field, the $\mathbf{Z_{Cl}}$-axis passes through the MC's axis and is aligned with the projection of the trajectory of the spacecraft (relative to the MC's velocity, which is approximately along the $\mathbf{X_{GSE}}$-axis) onto the cross-section of the MC, and $\mathbf{Y_{Cl}} = \mathbf{Z_{Cl}} \times \mathbf{X_{Cl}}$. See Figure 3 which shows the circular cross-section of an ideal MC and the projection of the spacecraft's path onto the cross-section in Cl coordinates, and see the *Wind*/MFI Website https://wind.gsfc.nasa.gov/mfi/ecliptic.html for further discussion and derivation of the coordinate transformation from GSE coordinates to Cl coordinates.

**Appendix B      Criteria for Estimating Quality of Magnetic Cloud Fit**

Here we quantify the quality ($Q_0$) of the model parameter-fit of a given magnetic cloud (MC) into three possibilities, $Q_0 = 1,2,3$, for excellent, good/fair, and poor, respectively, given below in terms of magnetic field quantities resulting from use of the MC model (Lepping *et al.*, 1990). However, for the sake of compactness we often refer to Quality as a measure of the MC *per se*, where it is mainly the quality of the MC parameter-fit that is being estimated.

We first describe the characteristics of those MC's that fall into the $Q_0 = 3$ (poor) category. This category arises from satisfying any one of the following $Q_0 = 3$ criteria: |check| ≥55%, |CA| ≥97%, $<B_X>_{Cl}$ ≤ -1.5 nT, either the f-flag or the F-flag = NOT OK, Diameter ≥ 0.45 AU, asf ≥ 40%, Cone angle ($\beta_{CA}$) ≤ 25° or $\beta_{CA}$ ≥ 155°, and $\chi_R$ ≥ 0.215. Notice that $\chi_R$ = 0.215 corresponds to a MC field noise level ν of 4.0 nT, according to Lepping *et al.* (2003, 2004), and this is the highest MC noise level that they found acceptable. The remaining cases, comprising designated "$Q_0 = 1$ or 2," are next examined to differentiate the "good" cases ($Q_0 = 1$) from the "fair" ($Q_0 = 2$) ones. The $Q_0 = 1$ cases must satisfy all of the following criteria: |check| ≤ 20%, $|<B_Y>_{Cl}|$ ≤ 3.0 nT, asf ≤ 30%, 45° ≤ $\beta_{CA}$ ≤ 135°, and $\chi_R$ ≤ 0.165. These are the "$Q_0 = 1$ set." Notice that $\chi_R$ = 0.165 corresponds to a MC field noise level ν of 3.0 nT, according to Lepping et al. (2003a; 2004). The remaining cases within set 1,2, *i.e.*, those not satisfying the $Q_0 = 1$ criteria, are put into category $Q_0 = 2$.

There are many ways that a MC can achieve a $Q_0 = 3$ quality, so there is no typical $Q_0 = 3$ MC. However $\chi_R$ and asf are usually the two most important parameters in judging MC quality. The quality criteria (meaning for all $Q_0 = 1,2,3$) were derived from our experience in the application of the Lepping *et al.* (1990) model and partly from a desire to be consistent with the results of the error study by Lepping *et al.* (2003, 2004). It should be stressed that, by our criteria, a MC may well satisfy the original Burlaga *et al.* (1981) definition of a MC and still not have good flux rope structure by the Lepping *et al.* (1990) model and therefore not qualify for a $Q_0$ of 1 or 2.